\documentclass[journal,twocolumn]{IEEEtran}

\usepackage{multirow,cite,algorithm,algorithmic,amsmath,amssymb,epsfig,mathrsfs,subfigure,url,color,epstopdf}
\renewcommand{\baselinestretch}{1.0}

\usepackage{graphicx}
\usepackage{tabularx}

\usepackage{cancel}

\newcommand{\beq}{\begin{equation}}
\newcommand{\eeq}{\end{equation}}
\newcommand{\beqn}{\begin{eqnarray}}
\newcommand{\eeqn}{\end{eqnarray}}

\newcommand{\qed}{\nobreak \ifvmode \relax \else
      \ifdim\lastskip<1.5em \hskip-\lastskip
      \hskip1.5em plus0em minus0.5em \fi \nobreak
      \vrule height0.75em width0.5em depth0.25em\fi}
						\usepackage[margin=0.7 in]{geometry}

      \usepackage[table,xcdraw]{xcolor}

\definecolor{lightblue}{rgb}{0.83,0.85,1.0}
\definecolor{white}{rgb}{1.0,1.0,1.0}
\definecolor{light-gray}{gray}{0.95}
\definecolor{BLUE}{rgb}{0,0,1}

\begin{document}

\title{Handover Management in 5G and Beyond: A Topology Aware Skipping Approach}

\author{Rabe~Arshad,
        Hesham~ElSawy,
        Sameh~Sorour,
        Tareq~Y.~Al-Naffouri,
        and~Mohamed-Slim~Alouini
\thanks{Rabe Arshad is with the Department
of Electrical Engineering, King Fahd University of Petroleum and Minerals, Saudi Arabia. E-mail: g201408420@kfupm.edu.sa} 
\thanks{Hesham Elsawy, Tareq Y. Al-Naffouri and Mohamed-Slim Alouini are with the CEMSE Division, EE Program, King Abdullah University of Science and Technology, Saudi Arabia. Email: \{hesham.elsawy, tareq.alnaffouri, slim.alouini\}@kaust.edu.sa}
\thanks{Sameh Sorour is with the Department of Electrical and Computer Engineering, University of Idaho, USA. E-mail: samehsorour@uidaho.edu}
}


\maketitle

\begin{abstract}
Network densification is found to be a potential solution to meet 5G capacity standards. Network densification offers more capacity by shrinking base stations' (BSs) footprints, thus reduces the number of users served by each BS. However, the gains in the capacity are achieved at the expense of increased handover (HO) rates. Hence, HO rate is a key performance limiting factor that should be carefully considered in densification planning. This paper sheds light on the HO problem that appears in dense 5G networks and proposes an effective solution via topology aware HO skipping. Different skipping techniques are considered and compared with the conventional best connected scheme. To this end, the effectiveness of the proposed schemes is validated by studying the average user rate in the downlink single-tier and two-tier cellular networks, which are modeled using Poisson point process and Poisson cluster process, respectively. The proposed skipping schemes show up to $47\%$ gains in the average throughput that would maximize the benefit of network densification.


\end{abstract}


\IEEEpeerreviewmaketitle

\section{Introduction}

\IEEEPARstart{D}{ue} to the rapid proliferation of mobile phones, tablets, and other handheld devices, an increasing traffic demand is observed worldwide. This drastically increasing capacity demand has driven the evolution of cellular networks from macro base stations (BSs) deployment to small, pico, and even nano BSs. For instance, the 5G evolution for cellular networks dictates 1000 fold capacity improvement, which is expected to be fulfilled by an evolutionary heterogeneous network densification phase~\cite{1}. Deploying more BSs within the same geographical region shrinks the footprint of each BS, which increases the spatial spectral efficiency and offers more capacity. However, the foreseen capacity gains offered by network densification are achieved at the expense of increased handover (HO) rates. Such important negative impact of BS densification (i.e., HO rate) is usually overlooked~\cite{5G}. In addition to the HO signaling overhead, the HO procedure interrupts the data flow to the user due to link termination with the serving BS and link establishment with the target BS. Increasing the HO rate increases the frequency of such undesirable interruptions as well as the associated signaling overhead, which may diminish or can even nullify the foreseen network densification capacity gains. Consequently, any discussion about network densification is never complete without incorporating the corresponding HO cost.\\

The HO process by itself is a core element of cellular networks to support user mobility. Hence, HO management has always been a focal research point in the context of cellular networks (see \cite{trends} for GSM/CDMA and \cite{LTEsurvey2,HOsurvey} for LTE systems). Modeling and improving the handover performance has been extensively addressed in the cellular network literature. For instance, the cell dwell time is characterized in~\cite{celldwell} for the circular and hexagonal shaped cells. An analytical model, based on application-specific signal strength tuning mechanism is presented in~\cite{vertical-optimization} to help optimizing the vertical HOs. HO signaling overhead reduction algorithms are proposed in~\cite{zhangsignalling} for two tier networks and in~\cite{zhangCRAN} for cloud-RAN based heterogeneous networks. A HO management technique, based on self organizing maps is proposed in~\cite{kernel} to reduce unnecessary HOs for indoor users in two tier cellular networks. The authors in~\cite{HOdelay1} present a study to avoid unnecessary vertical HOs and reduce the overall packet delay for low speed users in two tier downlink cellular networks. HO delay for a single tier network is characterized in~\cite{HOdelayrelation}. However, none of the aforementioned studies tackle the interplay between HO cost and capacity gain as a function of the BS intensity.\\ 

\begin{figure}[!t]
\centering
\hspace{-0.5cm}
\includegraphics[width=1 \linewidth]{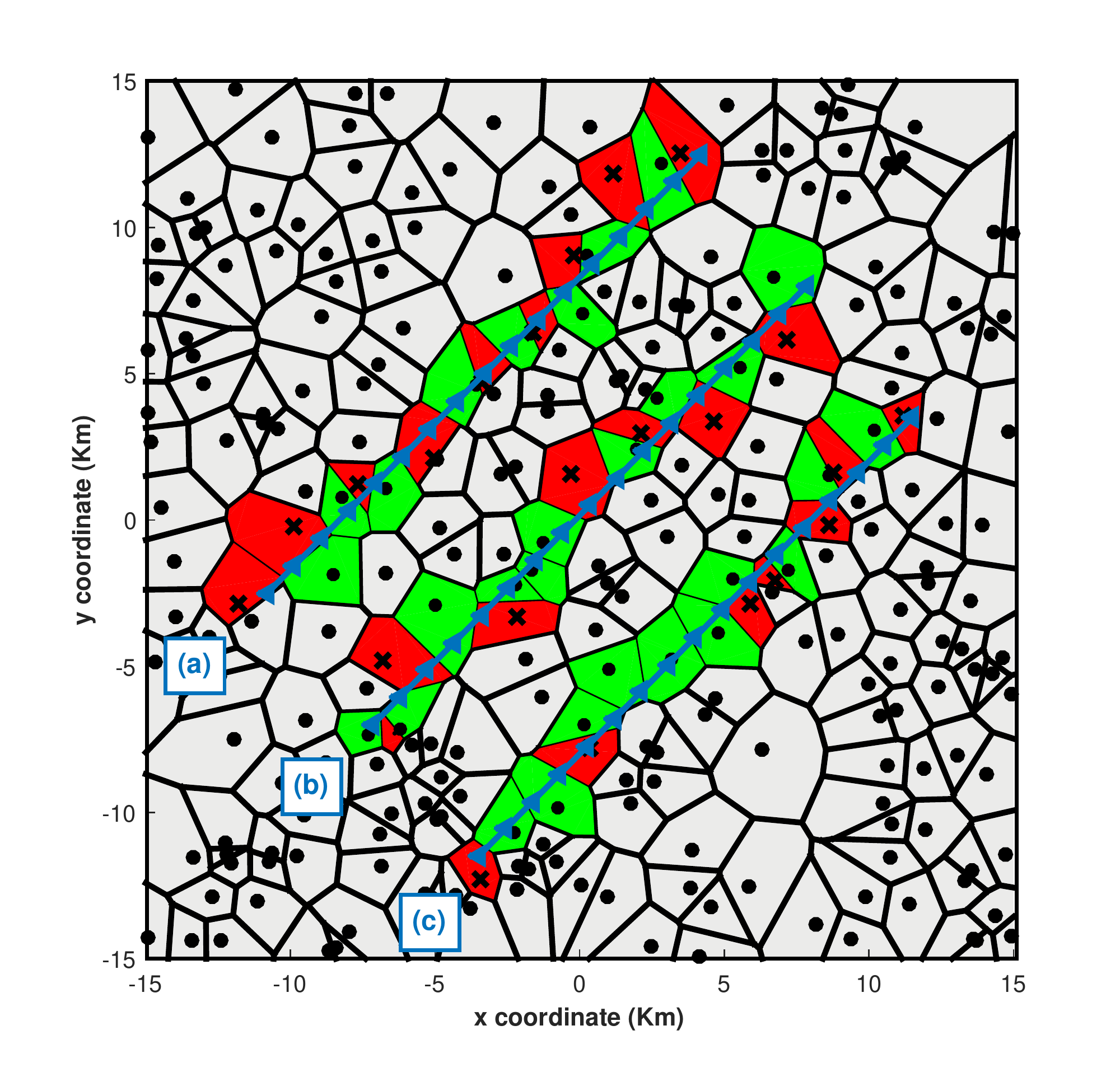}
\small \caption{Voronoi tessellation of a PPP based single tier cellular network with black circles representing the BSs' locations in 30 km x 30 km grid. (a), (b), and (c) represent alternating, user location aware, and cell-size aware HO skipping schemes, respectively. Blue line represents user trajectory while green and red colors denote serving (circles) and skipped (cross) BSs' coverage areas, respectively.}
\label{model}
\end{figure}

Stochastic geometry, which is a widely accepted mathematical tool to model and analyze cellular networks, enables performance characterization in terms of the BS intensity as well as other physical layer parameters (see~\cite{6a} for a survey and~\cite{tutorial} for a tutorial). For cellular networks modeled via the Poisson point process (PPP),~\cite{jeff_rate} studies the throughput gains as a function of the BS intensity for static users. Such model is extended to the case of Poisson cluster process (PCP) in \cite{Ghrayeb}. The HO rate for PPP cellular networks in terms of the BS intensity is characterized in~\cite{Lin} for single-tier scenario and in~\cite{10a} for multi-tier scenario. The work in~\cite{10a} is extended to the case of PCP in~\cite{Hocluster}. The HO rate for the recently proposed Phantom cells is characterized in \cite{macro-assisted}. However, none of the aforementioned studies investigate the integrated effect of network densification (i.e., BS intensity) in terms of both the HO cost and the throughput gains. The HO negative impact on the average throughput is studied in~\cite{sadr2015handoff, zhangdelay, ge2015user}. However, none of~\cite{sadr2015handoff, zhangdelay, ge2015user} proposed a solution for the HO problem. The authors in~\cite{cu-split} proposed a control/data plane split architecture with macro BS anchoring to mitigate the HO effect on user throughput. However, the proposed solution in~\cite{cu-split} necessitates a massive network upgrade. The authors in~\cite{icc, globe} propose a simple HO skipping scheme that is compatible with the current cellular architecture to mitigate the HO cost in a single tier cellular network. Particularly,~\cite{icc, globe} advocate sacrificing the best BS association and skip some HOs along the user trajectory at high speeds to reduce the number of handovers. Such skipping strategy has shown a potential to improve the user throughput at high velocities despite sacrificing the always best connectivity strategy. The skipping scheme in~\cite{icc, globe} is extended to two-tier networks in~\cite{velocityaware}. However, the HO skipping schemes presented in~\cite{icc,globe,velocityaware} are topology agnostic, which may result in non-efficient skipping decisions. Particularly,~\cite{icc,globe,velocityaware} advocate an alternate HO skipping approach in which the user skips associating to every other BS along its trajectory irrespective of the cell-size (coverage area) and/or the path of the trajectory through the cell as shown in Fig.~\ref{model}(scheme a). Consequently, there could be cases where the dwell time inside the cell of a skipped BS is larger than the dwell time inside a non-skipped BS. Articulated differently, the user may skip necessary HOs to BSs that cover a large sections of the user trajectory. To this end, devising smarter HO skipping schemes still entails to be inscribed for future cellular networks.\\


In this paper, we exploit topology awareness and user trajectory estimation to propose smart HO management schemes in single and two tier downlink cellular networks.\footnote{We assume that the network has the information about the user trajectory within the target BS footprint. In some cases like users riding monorails in downtowns, the user trajectory across the target BS footprint is fixed and known. For other cases, several studies including~\cite{trajectory1,trajectory} are conducted in the literature on the estimation of mobile user trajectory.} The proposed schemes account for the location of the trajectory within the cells and/or the cell-size to take the HO skipping decision. The performance of the proposed schemes is analyzed using tools from stochastic geometry. More specifically, we consider two network scenarios, namely, a single-tier cellular network abstracted by PPP and a two tier cellular network abstracted by PPP macro BSs overlaid with PCP small BSs. Then we study the impact of HO delay on user throughput in the two network scenarios and show the gains and effectiveness of the proposed schemes by Monte Carlo simulations. The results manifest the HO problem of the always best connectivity scheme at high speeds in dense cellular environments. The user throughput via the proposed skipping schemes outperforms the always best connected scheme at velocities starting from 30 km/h. Particularly, for BSs intensity of 50 BS/km$^2$, the proposed schemes show up to $8\%$ more rate gains with respect to (w.r.t.) best connectivity and $23\%$ w.r.t. alternating skipping over the user velocity of 100 km/h, which is the average monorail speed in downtown. More gains are expected at higher BSs intensities. Finally, several insights into the design of HO skipping and the effective range of velocities for each of the proposed skipping schemes are presented.\\

The rest of paper is organized as follows. Section II overviews the HO procedure and presents the proposed HO schemes. Section III analyzes the performance metrics (e.g. coverage probability, HO cost, and average throughput) for proposed HO skipping schemes in single tier network. Section IV shows the significance of proposed model in two-tier networks. Finally, the paper is concluded in Section V.

\section{Overview of Handover Process}

HO is the process of changing the user equipment (UE) association with mobility such that the best serving BS is always selected. One popular and simple rule for determining the best serving BS is based on the received signal strength (RSS) level. That is, the UE changes its association if another BS provides a higher RSS than the serving BS, which may happen when the user moves away from the serving BS towards another BS. With the increasing heterogeneity of cellular networks, many other criteria are developed for selecting the best serving BS which may include load balancing, delay, and throughput metrics~\cite{alpha-load,MIMO,Hesham-traffic}. Despite of the selection rule, the UE will always change its association with mobility and the HO rate increases with the BS density. Hence, the HO cost is always an increasing function of the BS density.\\

In general, HO is performed in three phases: initiation, preparation, and execution. During the initiation phase, the user reports reference signals measurements from neighboring BSs to the serving BS. For instance, the signal measurement report in 4G Long Term Evolution (LTE) includes, but not limited to, reference signal received power (RSRP) and reference signal received quality (RSRQ) (see~\cite{LTEHO} for the HO procedure in LTE). Also, the HO may be initiated based on downlink and/or uplink signal measurement reports. In the next phase, which is the preparation phase, signaling is exchanged between the serving BS, target BS, and the admission controller. The admission controller makes a decision about the initiation of the HO based on network defined HO criteria. Once the HO criteria are met, the user releases the serving BS and attempts to synchronize and access the target BS using the random access channel (RACH). Upon synchronization with the target BS, the UE sends the confirmation message to notify the network that the HO is executed. The aforementioned HO procedure involves signaling overhead between the user, serving BS, target BS, and core network, which interrupts the data flow and decreases the throughput of mobile user. The frequency at which such interruptions happen is a function of the relative values of the BS intensity and user velocity. The duration of each interruption, denoted as HO delay, measured from the beginning of initiation phase to the end of execution phase can be significant~\cite{14a}. Consequently, at high velocities and/or dense cellular environment, it is desirable to decrease the frequency of such HO interruptions, which motivates the HO skipping scheme. Note that high mobility can exist in dense cellular environments such as riding monorails or driving over elevated highways that go through downtowns.\\

HO skipping reduces the frequency at which the HO process is performed by sacrificing some of the best BS connectivity associations. Hence, maintaining longer service durations with the serving BSs and reducing HO delay.
For instance, in an RSS based association scheme with universal frequency reuse, HO skipping sacrifices some best signal-to-noise-plus-interference-ratio (SINR) associations along the trajectory. When the user skips the HO to the BS providing the strongest signal, denoted as blackout (BK) phase, the interference from the skipped BS may be overwhelming to the SINR. To improve the SINR during blackout, nearest BS interference cancellation (IC)~\cite{16a} and non-coherent cooperative BS service via coordinated multipoint (CoMP) transmission~\cite{3gpp, crancomp1, crancomp2} can be exploited to improve the SINR during blackout.\footnote{Non-coherent CoMP is used as channel state information is hard to predict at high velocities.} In the cooperative BS service, the user can be jointly served by the serving BS and the next target BS. In addition to IC and CoMP, the performance of the skipping scheme can be further improved by reducing the blackout durations along the users trajectories via smart skipping schemes. Different from the topology agnostic (i.e., alternating skipping) approach proposed in \cite{icc,globe,velocityaware}, this paper focuses on the following three novel variants of HO skipping. Note that all of the following skipping schemes assume that the trajectory within the target BS footprint is known via some trajectory estimation techniques~\cite{trajectory1,trajectory}.

\subsubsection{\bf Location-Aware HO Skipping}
 The location aware HO skipping scheme accounts for the shortest distance between the user trajectory and the target BS to decide the HO skipping. That is, the HO skips associating to the target BS if and only if the minimum distance along the user trajectory and the target BS exceeds a pre-defined threshold $L$. The threshold $L$ can be designed such that the user skips the BSs in which the trajectory passes through the cell edge only. The location aware HO skipping scheme is illustrated in Fig.~\ref{model}(scheme b).\\

\subsubsection{\bf Cell-Size Aware HO Skipping}
 Cell-size aware HO skipping scheme allows users to skip HOs to target BSs that have a footprint (i.e., service area) less than a pre-defined threshold $s$. Since the cell dwell time depends of the BS footprint size, size aware skipping scheme aims at avoiding large blackout durations. Hence, it allows users to skip small sized cells and associate with large cells. The cell-size aware HO skipping scheme is illustrated in Fig.~\ref{model}(scheme c). Note that it is implicitly assumed that the service areas of all BSs are known to the network, which can be inferred from several network planning tools used by cellular operators such as Aircom Asset~\cite{asset} and Mentum Planet~\cite{mentum}. Such tools take antenna characteristics, BS configuration, terrain and clutter information into account to predict cell-sizes.\\

\subsubsection{\bf Hybrid HO Skipping}
Neither the location aware skipping nor the size aware skipping alone accurately reflects the true cell dwell time. Hence, combining both schemes gives a better inference about the cell dwell time, which can improve the HO skipping decisions and performance. Consequently, the hybrid HO skipping scheme combines both location awareness and cell-size awareness to decide which BS to skip. That is, it takes user location and cell area into account while making the decision for HO.\\

\section{Handover Skipping in Single Tier Networks}
In this section, we consider a single tier downlink cellular network, where the BSs' locations are modeled via a two-dimensional homogenous PPP $\Phi$ of intensity $\lambda$. It is assumed that all BSs transmit with the same power $P$. A general path loss model with path loss exponent $\eta>2$ is assumed. Without loss of generality, we focus on a test mobile user and index all BSs with an ascending order according to their instantaneous distances from the test user. By Slivnyak-Mecke theorem for the PPP~\cite{20a}, the performance of any other user in the network is equivalent to the performance of the test user. Defining $R_{k}$ as the distance from the test user to the $k^{th}$ BS, then the inequalities $(R_1 < R_2 < R_3 <....)$ always hold. Channel gains are assumed to have $i.i.d.$ Rayleigh distributions with unit variance i.e., $h\sim \exp(1)$. We consider a universal frequency reuse scheme and study the performance of one frequency channel. We consider user mobility with constant velocity $v$ over an arbitrary long trajectory and assume that a HO is triggered when the user enters the voronoi cell of the target BS. We first analyze the coverage probability for all HO skipping cases and then evaluate the HO cost and average throughput with the simulation parameters shown in Table~\ref{tab1}.

\begin{table}[!t]
\caption{\: Simulation parameters for PPP based cellular network}
\center
\vspace{-0.5cm}
\resizebox{0.5\textwidth}{!}{
\begin{tabular}{|c c c c|}
\hline
\rowcolor{cyan}
\multirow{-1}{*}{\textcolor{white}{\textbf{Parameter}}} & \multirow{-1}{*}{\textcolor{white}{\textbf{Value}} } &\multirow{-1}{*}{ \textcolor{white}{ \textbf{Parameter}}}  & \multirow{-1}{*}{\textcolor{white}{ \textbf{Value }}}  \\ \hline  \hline
& & & \\
 \multirow{-2}{*}{Overall Bandwidth $W$:}             &  \multirow{-2}{*}{10 MHz }    &  \multirow{-2}{*}{ Path loss exponent $\eta$: }  &   \multirow{-2}{*}{4}     \\
 \cellcolor{cyan!20!}  & \cellcolor{cyan!20!}    &\cellcolor{cyan!20!}    & \cellcolor{cyan!20!}  \\
\multirow{-2}{*}{\cellcolor{cyan!20!} BS intensity $\lambda$:}  & \multirow{-2}{*}{\cellcolor{cyan!20!}50 BS/km$^{2}$} & \multirow{-2}{*}{\cellcolor{cyan!20!}HO delay $d$:}  &   \multirow{-2}{*}{\cellcolor{cyan!20!} 1 s}   \\
& & & \\
 \multirow{-2}{*}{Size Threshold $s$:}               & \multirow{-2}{*}{ 1.28/$\lambda$ km$^2$}  &  \multirow{-2}{*}{Location Threshold $L$:}       &  \multirow{-2}{*}{ 2.3/$\lambda$ m}   \\
\cellcolor{cyan!20!} & \cellcolor{cyan!20!}  & \cellcolor{cyan!20!}  &\cellcolor{cyan!20!}   \\
 \multirow{-2}{*}{\cellcolor{cyan!20!} Hybrid Thresholds $s$, $L$:}           &  \multirow{-2}{*}{\cellcolor{cyan!20!} 0.38/$\lambda$ km$^2$, 1.8/$\lambda$ m }     & \multirow{-2}{*}{\cellcolor{cyan!20!} Tx Power $P$:}     &  \multirow{-2}{*}{\cellcolor{cyan!20!} 1 watt} \\  \hline
\end{tabular}
}
\vspace{3mm}
\label{tab1}
\end{table}


\subsection{Coverage Probability}
The coverage probability is defined as the probability that the received SINR by the test user exceeds a certain threshold $T$. The coverage probability for the best connected case is given by

\begin{eqnarray}
\mathcal{C}_{BC} &=& \mathbb{P} \left\{ \frac{P h_1  R_{1}^{-\eta}}{ \sum_{i\epsilon \Phi \backslash b_1}{}P h_{i}  R_{i}^{-\eta} + \sigma^2} > T \right\}
\label{C1}
\end{eqnarray}

\noindent where the nearest BS, denoted as $b_1$, is removed from the interfering BSs in \eqref{C1} because the serving BS do not contribute to the aggregate interference.

In the blackout case, the test user is not served from the nearest BS due to HO skipping. Instead, the test user maintains the association with the serving BS (which is not the nearest anymore) or handovers the connection to the next target BS depending on their relative distances during blackout. If CoMP is enabled, then the test user is simultaneously served by both the serving and the next target BSs during the blackout phase. Let $R_s$ and $R_t$ denote the distances from the test user to the serving BS (denoted as $b_s$) and next target BS (denoted as $b_t$) during the blackout phase. Then the coverage probabilities for the blackout case with IC capabilities without and with BS cooperation are given by $\mathcal{C}^{(1)}_{BK(IC)}$ and $\mathcal{C}^{(2)}_{BK(IC)}$, respectively.

\begin{eqnarray}
\hspace{0.4cm} \mathcal{C}^{(1)}_{BK(IC)}=\mathbb{P}\left\{\frac{P h_x \min(R_{s},R_t)^{-\eta}}{\sum_{i\epsilon\Phi\backslash b_{1},b_{x}}^{} Ph_{i} R_{i}^{-\eta}+\sigma^2} > T  \right\}
\label{C2}
\end{eqnarray}

\noindent where the subscript $ x=s$ if $R_s<R_t$ and $ x=t$ otherwise.

\begin{eqnarray}
\mathcal{C}^{(2)}_{BK(IC)}=\mathbb{P} \left\{\frac{|\sqrt{P}g_{x}R_{s}^{-\eta/2}+\sqrt{P}g_{t} R_{t}^{-\eta/2}|^2}{\sum_{i\epsilon\Phi\backslash b_{1},b_{s},b_{t}}^{} Ph_{i} R_{i}^{-\eta}+\sigma^2}>T\right\}
\label{C3}
\end{eqnarray}
where $g_x$ and $g_t$ are zero-mean and unit-variance complex Gaussian channels to reflect the non-coherent CoMP transmission. Note that $b_1$ in~\eqref{C2} and~\eqref{C3} is the skipped BS whose signal power is eliminated from the aggregate interference by virtue of IC.

The coverage probability for the best connected scenario given in equation \eqref{C1} is mathematically characterized in~\cite{7a}. Furthermore, the coverage probability for the HO skipping (i.e., blackout) scenarios given in \eqref{C2} and \eqref{C3} are mathematically characterized in \cite{icc} and \cite{globe}. However, it is highly difficult to conduct tractable analysis for the proposed HO skipping schemes due to the random shape of the voronoi cell and the random location and orientation of the trajectory within the voronoi cell. Therefore, we show the coverage probabilities for the best connected and HO skipping cases based on Monte Carlo simulations. The simulations in this paper follow~\cite{icc,globe,velocityaware} where both the mathematical analysis and the simulations are used and validated.

\begin{figure}[!t]
\centering
\includegraphics[width=0.7 \linewidth]{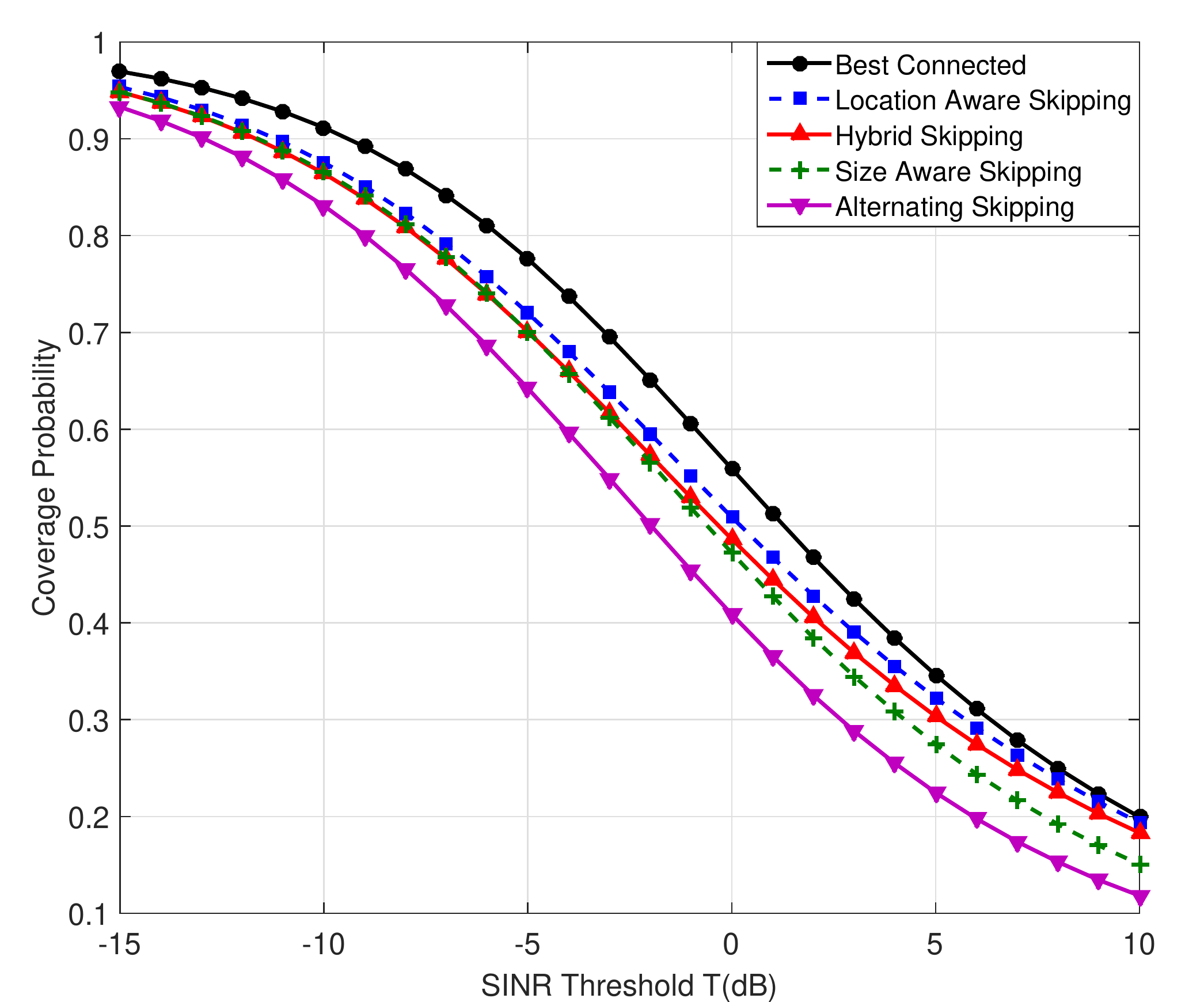}
\small \caption{Coverage probability vs. SINR threshold for best connected and HO skipping cases.}
\label{CP1}
\end{figure}

\begin{figure}[!t]
\centering
\includegraphics[width=0.7 \linewidth]{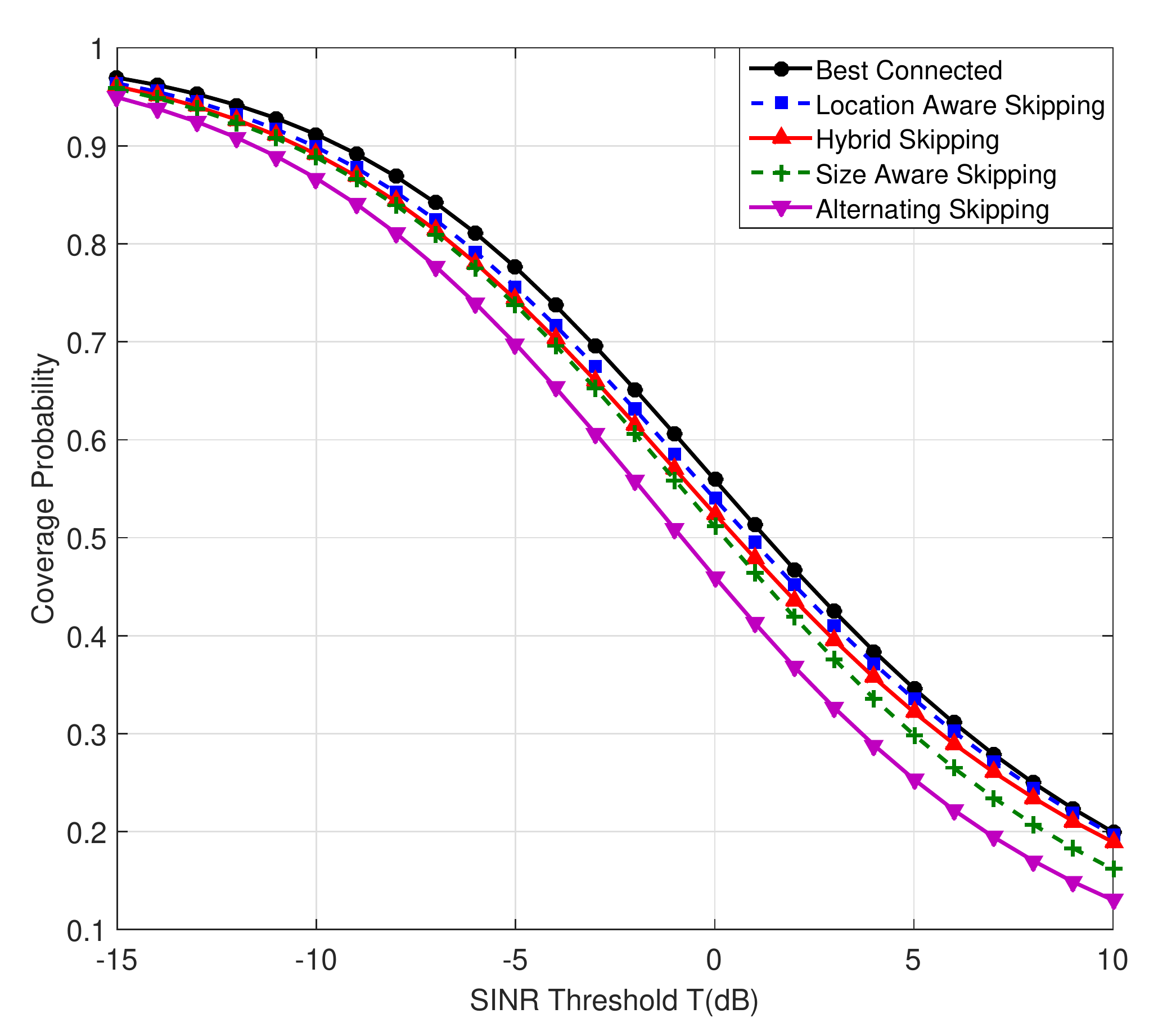}
\small \caption{Coverage probability vs. SINR threshold for best connected and HO skipping cases with CoMP transmission.}
\label{CP2}
\end{figure}

 Figs.~\ref{CP1} and \ref{CP2} show the coverage probability plots for the best connected and HO skipping cases without and with BS cooperation, respectively. As expected, sacrificing the always best connectivity reduces the average coverage probability over the user trajectory. Nevertheless, employing a smart skipping scheme via location and size awareness can mitigate such coverage probability reduction. Furthermore, comparing the results in Figs.~\ref{CP1} and \ref{CP2} quantifies the contribution of BS cooperation to the coverage probability. Note that the hybrid scheme shown in Figs.~\ref{CP1} and \ref{CP2} uses more relaxed size and distance constraints than the locations and size aware schemes as shown in Table I. Hence, it is able to have more HO skips with comparable coverage probability to the locations and size aware schemes. Note that the coverage probabilities shown in Figs.~\ref{CP1} and \ref{CP2} reflect the negative impact only of the HO skipping. The next section incorporates the HO cost into the analysis in order to fairly assess HO skipping.


\subsection{Handover Cost}
This section evaluates the HO cost for the best connected and HO skipping cases. The HO cost $\mathcal{D}$ is defined in terms of the normalized HO delay, which is given by

\begin{align}
\mathcal{D}= \min(\mathcal{H}_t \times d,1)
\end{align}

\noindent where $\mathcal{H}_t$ is the handover rate per unit time and $d$ is the delay in seconds per handover. Hence, the handover cost $\mathcal{D}$ is a unit-less quantity used to quantify the fraction of time wasted without useful data transmission along the user trajectory, which is due to handover signaling and radio link switching between serving and target BSs. Note that if $\mathcal{H}_t \times d \geq1$, this means that the cell dwell time is less than the handover delay. Hence, the entire time is wasted in handover signaling without useful data transmission and $\mathcal{D}$ is set to one.

The HO rate for a PPP based single tier network is characterized in~\cite{10a} for a generic trajectory and mobility model as
\begin{align}
\mathcal{H}_t= \frac{4v}{\pi}\sqrt{\lambda}
\end{align}
In order to calculate the HO rate via simulations, we first calculate the number of HOs per unit length and then multiply it with the user velocity. The number of handover per unit length is calculated by dividing the number of handovers by the trajectory length. Thus, $\mathcal{D}$ can be expressed as

\begin{align}
\mathcal{D}&= \mathcal{H}_l \times v \times d
\end{align}

\noindent where $\mathcal{H}_l$ is the number of HOs per unit length.

\begin{figure}[!t]
\centering
\includegraphics[width=0.7 \linewidth]{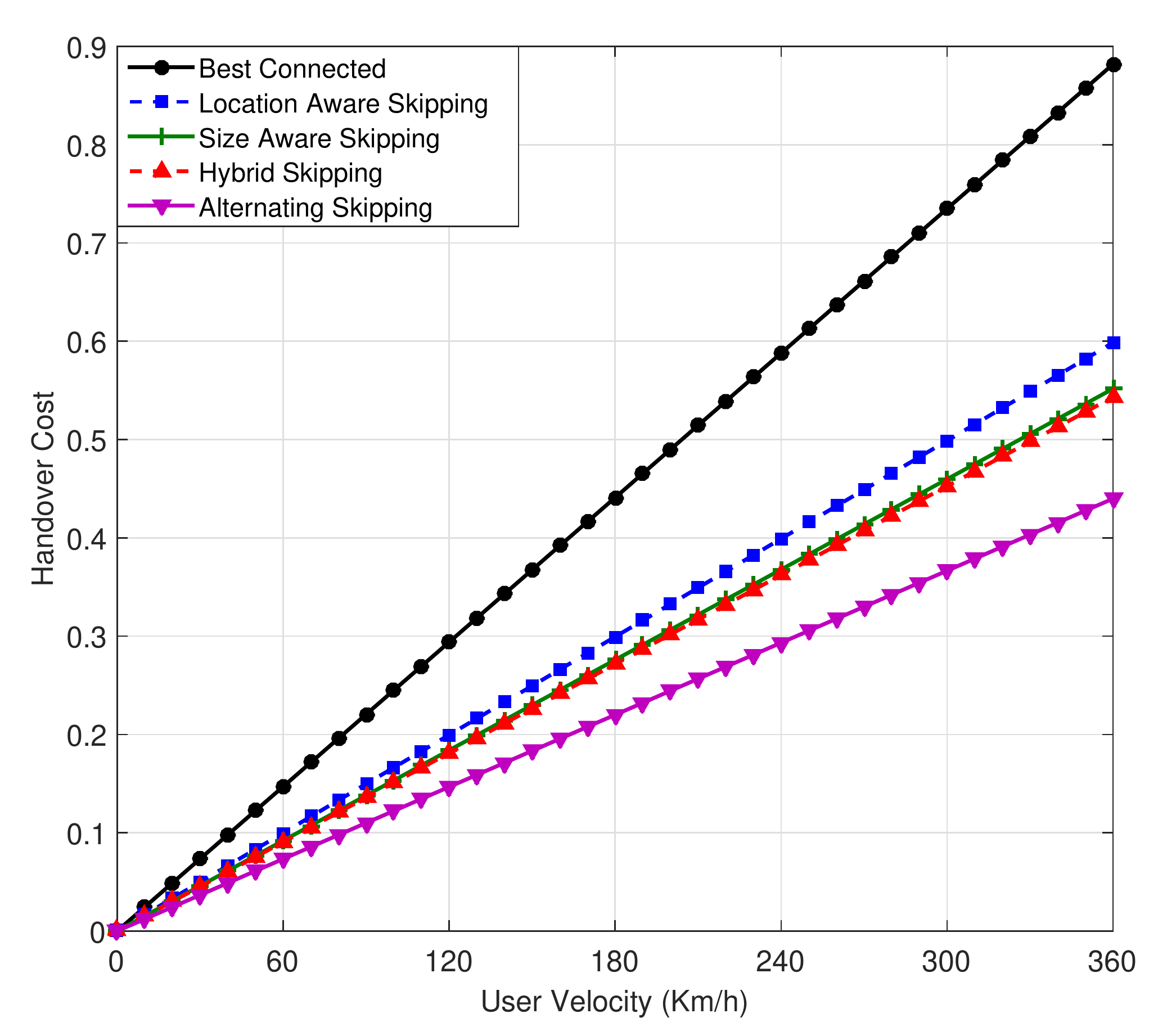}
\small \caption{Handover cost for conventional and HO skipping cases.}
\label{DHO}
\end{figure}

Fig~\ref{DHO} depicts the HO cost for the best connected and HO skipping schemes. Since the HO cost depends on the number of HOs, the best connectivity association shows significant HO cost as compared to the HO skipping strategies. The alternating HO skipping shows the minimal handover cost as it results in the maximum number of HO skips. However, the alternating skipping is topology agnostic and can have inefficient skipping decisions. Such inefficient decision can be reduced via location and size awareness on the expense of having higher HO cost (cf. Fig~\ref{DHO}) but better coverage probability (cf. Figs.~\ref{CP1} and \ref{CP2}). While Figs.~\ref{CP1} and \ref{CP2} focus on the negative impact of the skipping schemes, Fig~\ref{DHO} focuses on their positive impact. In the next section, the integrated effect (i.e., both the negative and positive) of the skipping schemes are assessed in the context of user throughput.


\subsection{Average Throughput}
Average throughput is a key performance indicator (KPI) for the cellular operators, which can be used to show the interplay between HO cost and capacity gain imposed by network densification. 
In this section, we quantify the effect of HO rate, and the impact of each of HO skipping schemes, on the average user throughput. The average throughput, denoted as $\mathcal{T}$, is defined as:

\begin{eqnarray}
\mathcal{T} &=& W\mathcal{R}(1-\mathcal{D}).
\label{through}
\end{eqnarray}
where $W$ denotes the overall bandwidth and $\mathcal{R}$ represents the ergodic spectral efficiency, which is defined by Shannon capacity formula as
\begin{eqnarray}
\label{rates}
\mathcal{R}&=&\mathbb{E}\big(\ln\big(1+{\rm SINR}\big)\big)
\end{eqnarray}

\begin{figure}[!t]
\centering
\includegraphics[width=0.7 \linewidth]{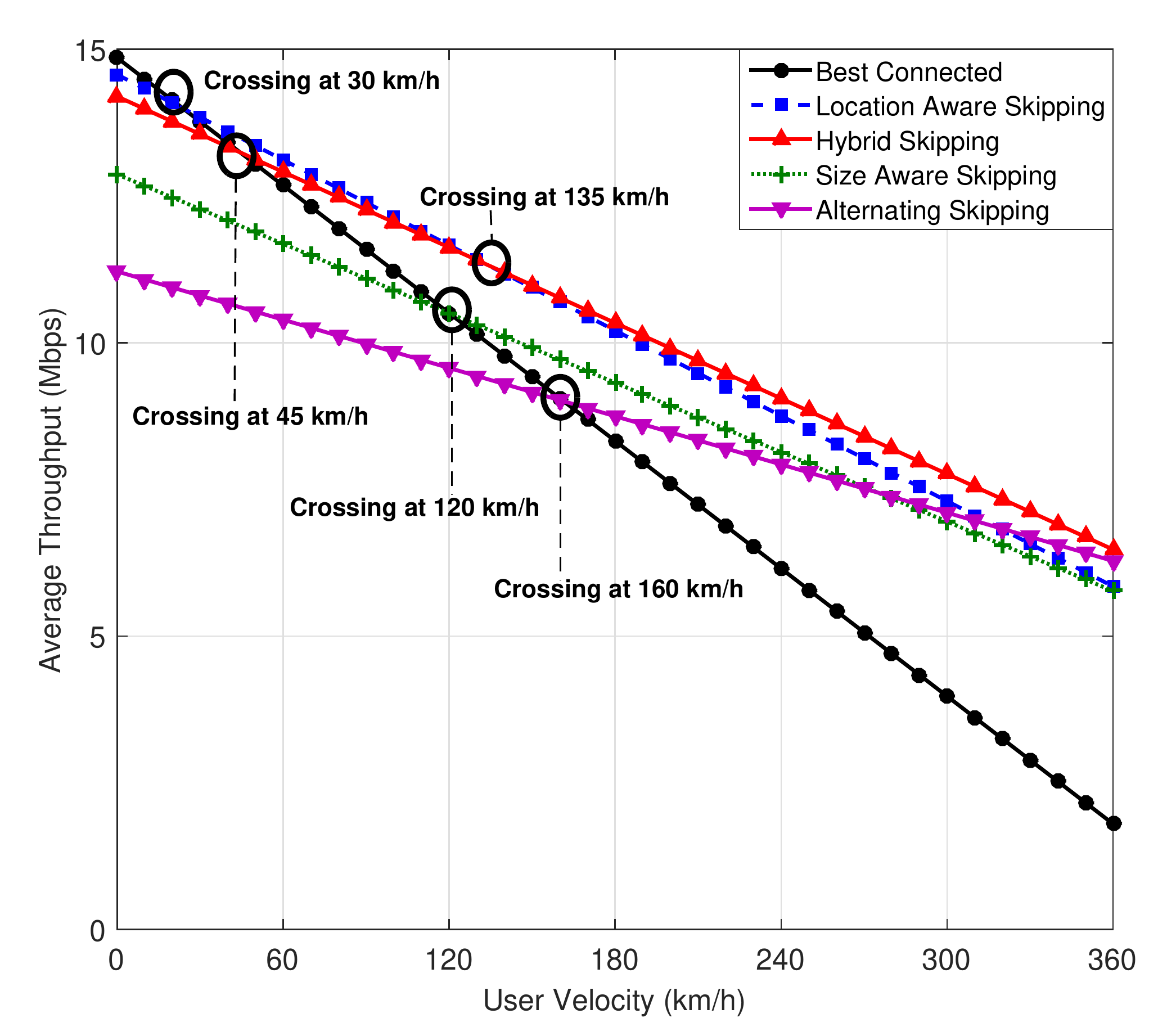}
\small \caption{Average throughput vs. user velocity for conventional and HO skipping cases.}
\label{th}
\end{figure}

\noindent Table~\ref{tab2} shows the spectral efficiencies for the best connected and HO skipping cases with and without IC capabilities, which are obtained via simulations using the definition in~\eqref{rates}. The spectral efficiencies given in Table~\ref{tab2} are used to obtain throughput plots via \eqref{through} as shown in Fig.~\ref{th}. The figure clearly shows the HO cost impact on the user throughput when the velocity increases. The figure also shows that the negative impact of the HO could be relieved using the skipping schemes. Particularly, the location aware HO skipping outperforms all other schemes at low velocities i.e. 30 km/h. With the proper adjustment of the hybrid skipping scheme, it outperforms the best connected association at 45 km/h and the location aware scheme at 135 km/h. Note that the slope of the hybrid curve is tunable through the cell-size threshold $s$ and the minimum distance threshold $L$. Size aware HO skipping is the least effective compared to the location aware and hybrid schemes, even though it shows considerable gains as compared to the topology agnostic (i.e., alternating HO skipping) scheme. Note that the size aware scheme is easier to implement than the location aware and hybrid schemes as it does not require complete information about the user trajectory in the target cell. Finally, the alternating HO skipping becomes comparable in performance with other schemes at very high user velocity (beyond 280 km/h) because the HO cost is significant and requires high number of skips to be mitigated.

\begin{table}[!t]
\renewcommand{\arraystretch}{1.05}
\caption{\: Spectral Efficiency for all cases in nats/sec/Hz}
\center
\resizebox{0.3\textwidth}{!}{
\begin{tabular}{|c c c|}
\hline
\rowcolor{cyan}
\multirow{-1}{*}{\textcolor{white}{\textbf{Scenario}}} & \multirow{-1}{*}{\textcolor{white}{\textbf{Non-IC}} } &\multirow{-1}{*}{ \textcolor{white}{ \textbf{IC}}} \\ \hline  \hline
& & \\
 \multirow{-2}{*}{Best connected ($\mathcal{R}_{BC}$)}             &  \multirow{-2}{*}{1.49}    &  \multirow{-2}{*}{-}      \\
\cellcolor{cyan!20!}  & \cellcolor{cyan!20!}    &\cellcolor{cyan!20!}  \\
\multirow{-2}{*}{\cellcolor{cyan!20!} Location Aware ($\mathcal{R}_{LA}$)}  & \multirow{-2}{*}{\cellcolor{cyan!20!}1.40} & \multirow{-2}{*}{\cellcolor{cyan!20!}1.45}  \\
& &  \\
 \multirow{-2}{*}{Hybrid ($\mathcal{R}_{HB}$)}               & \multirow{-2}{*}{1.36}  &  \multirow{-2}{*}{1.42}      \\
\cellcolor{cyan!20!}  & \cellcolor{cyan!20!}  & \cellcolor{cyan!20!}  \\
 \multirow{-2}{*}{ \cellcolor{cyan!20!} Size Aware ($\mathcal{R}_{SA}$) }           &  \multirow{-2}{*}{\cellcolor{cyan!20!} 1.21 }     & \multirow{-2}{*}{\cellcolor{cyan!20!} 1.28}     \\
  &  &  \\
 \multirow{-2}{*}{ Alternating ($\mathcal{R}_{AL}$)}           &  \multirow{-2}{*}{1.02 }     & \multirow{-2}{*}{1.11}     \\  \hline
\end{tabular}
}
\vspace{3mm}
\label{tab2}
\end{table}

\section{Handover Skipping in Two Tier Networks}

\begin{figure}[!t]
\centering
\includegraphics[width=0.7\linewidth]{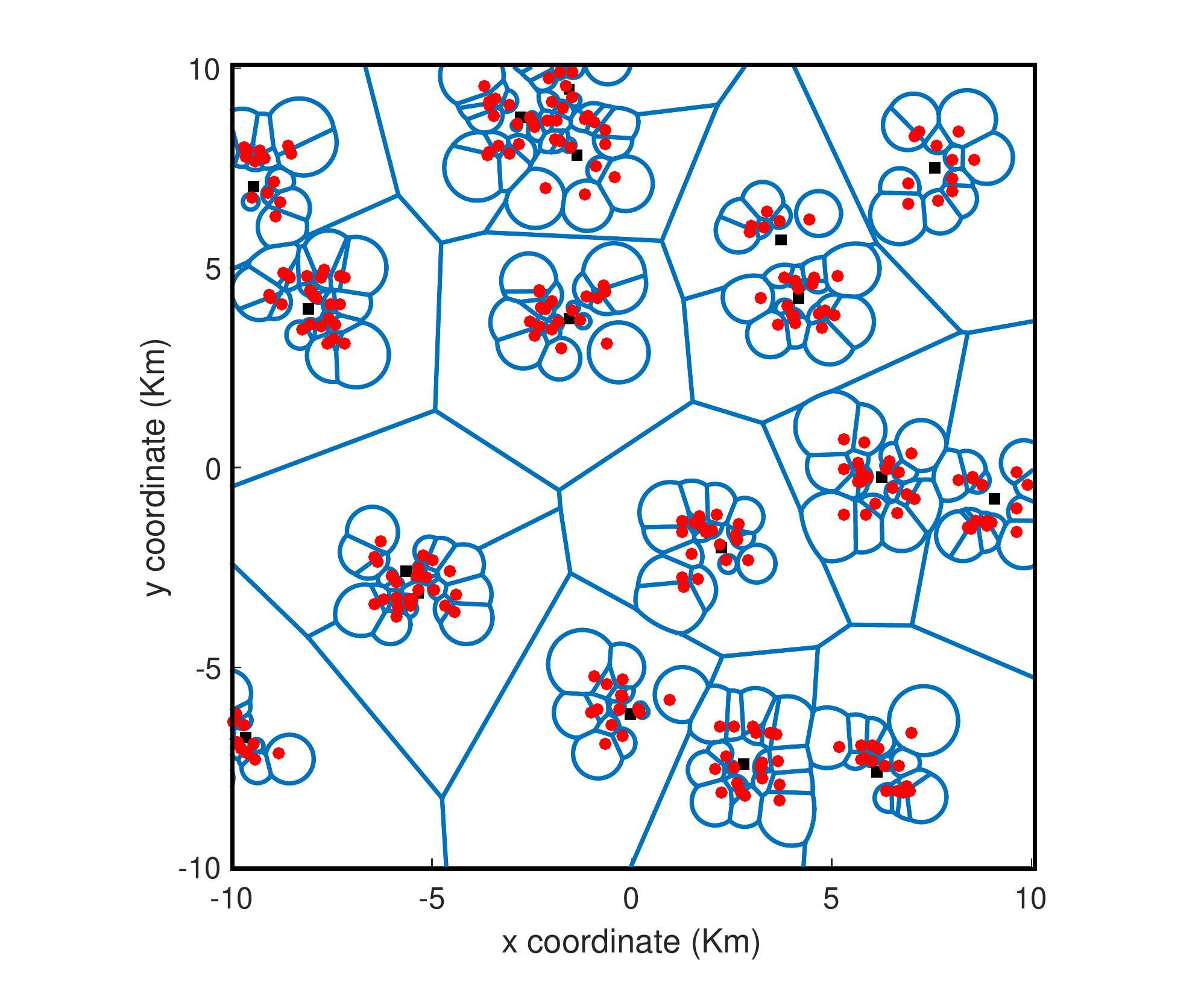}
\small \caption{Weighted Voronoi tessellation of two tier PCP based downlink cellular network with $\lambda_p= 0.04$ BS/km$^2$, $\lambda_c= 1$ BS/km$^2$, $P_1= 1$ watt, $P_2= 0.5P_1$ watt, $r= 2$ km. Black squares represent macro BSs while red circles denote femto BSs.}
\label{pcp}
\end{figure}

Current cellular networks are evolving towards a multi-tier architecture in which the macro BSs are overlaid with small BSs to cover traffic hotspots. Since hotspots are usually concentrated around popular/social regions, the small BSs are better modeled via a PCP~\cite{Ghrayeb}. The PCP is generated from a parent PPP in which each point in the parent PPP is replaced by multiple cluster points. The distribution of the cluster points around each parent point location determines the type of the PCP. In this work, we consider the Mat\'ern cluster process in which the parent points are generated via a homogenous PPP with intensity $\lambda_p$ while the daughter points are uniformly distributed within a ball of radius $r$, where the number of daughter points in each cluster follows poisson distribution with intensity $\lambda_c$. The parent points represent the macro BSs of tier-1 and the daughter points represent small BSs of tier-2 as shown in Fig.~\ref{pcp}. The total intensity of the BSs in the network becomes $\lambda^\prime= \lambda_{p}\lambda_{c}+\lambda_{p}$. It is assumed that the BSs belonging to the $i^{th}$ tier have same transmit power $P_{i}$, $i\in\{1,2\}$ and unity bias factor. A power-law path-loss model with path loss exponent $\eta_i>2$ is considered. Channel gains are assumed to have $i.i.d.$ Rayleigh distributions. Due to the different powers used by the macro and small BSs, the coverage regions in Fig.~\ref{pcp} are represented via a weighted voronoi tessellation~\cite{voronoi}.\\

For the considered two-tier network, we follow the same methodology in Section~III and study the users throughput to characterize the HO cost and assess the skipping solutions. We conduct our study on a test user moving with velocity $v$ and assume an RSS based association such that the HO is triggered when the user enters the voronoi cell of the target BS. Motivated by its superior performance when compared to all skipping schemes, this section focuses on the location aware skipping. Particularly, we compare the location aware skipping scheme for different distance thresholds to the always best connected strategy.

\begin{figure}[!t]
\centering
\includegraphics[width=0.7 \linewidth]{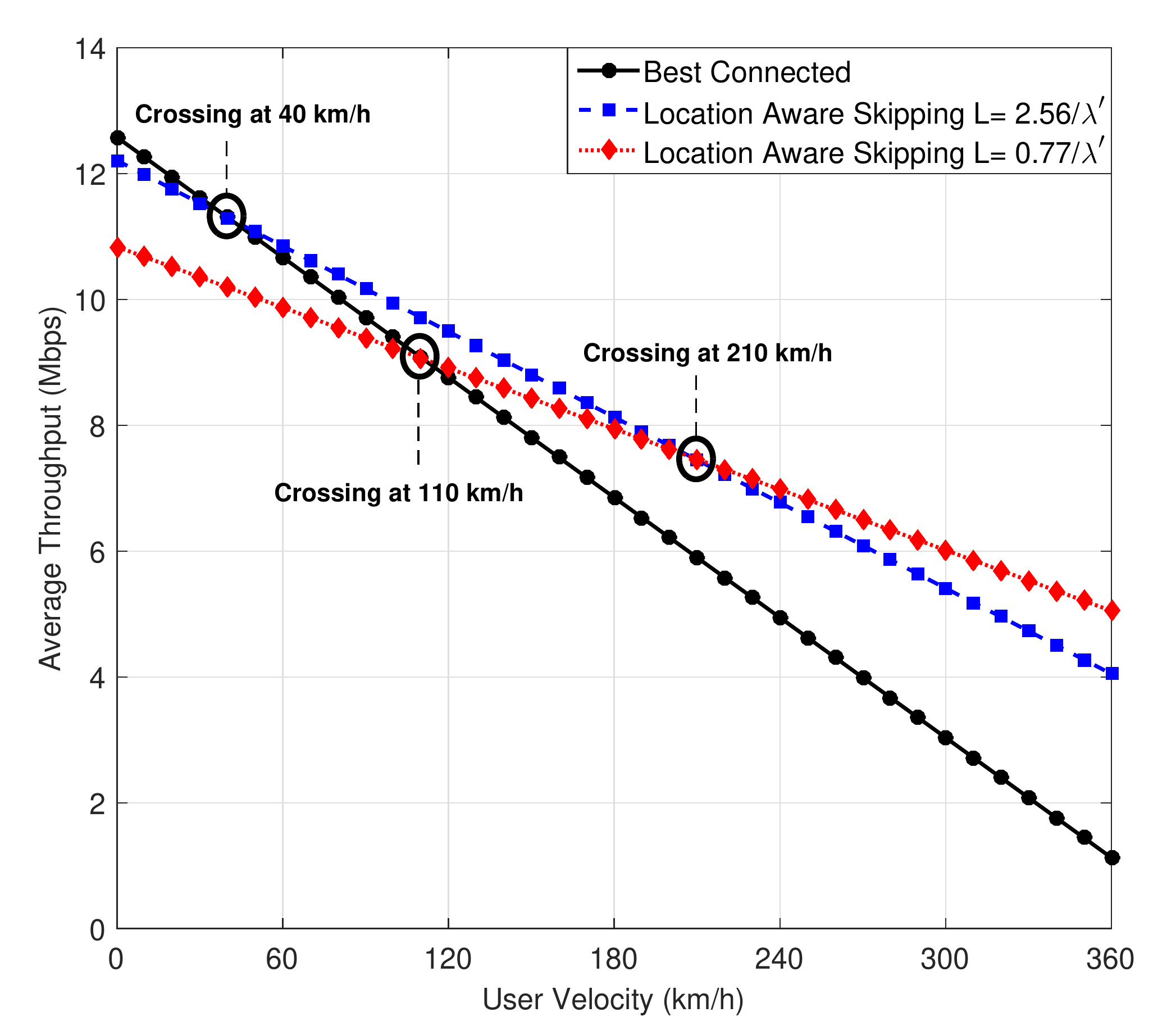}
\small \caption{Average throughput vs. user velocity for PCP based two tier network with $P_1= 1$ watt, $P_2= 0.1$ watt, $\lambda_p= 4$ BS/km$^2$, $\lambda_c= 12$ BS/km$^2$, $d= 1$ s, $r= 0.6$ km, $\eta_1=\eta_2=4$}
\label{thpcp}
\end{figure}

To assess the user throughput, we first evaluate the coverage probabilities, spectral efficiencies, and HO costs. Then the average throughput is calculated as in~\eqref{through}. Table~\ref{tab3} shows the spectral efficiencies for the best connected and location aware HO skipping schemes with distance threshold $L=0.77/\lambda^\prime$, $2.56/\lambda^\prime$. Fig.~\ref{thpcp} shows the average throughput plots for the best connected and location aware HO skipping cases. It is observed that the location aware HO skipping scheme in a PCP based cellular network outperforms the best connected association once the user exceeds 40 km/h. The results show up to $47\%$ throughput gains, which can be harvested through proposed smart handover strategy. From Fig.~\ref{thpcp}, it is observed that the location awareness with less threshold outperforms location awareness with higher distance threshold once the user exceeds 210 km/h. This is because decreasing the distance threshold $L$ relaxes the skipping constraint and increases the number of skips, which compensates for the excessive HO cost that happens at high mobility. It is worth noting that the considered clustering scheme in this paper is used for illustrative purposes only, in which similar results and insights apply to other clustering schemes.

\begin{table}[!t]
\renewcommand{\arraystretch}{1.05}
\caption{\: Spectral Efficiency for PCP Network in nats/sec/Hz}
\center
\resizebox{0.4\textwidth}{!}{
\begin{tabular}{|c c c|}
\hline
\rowcolor{cyan}
\multirow{-1}{*}{\textcolor{white}{\textbf{Scenario}}} & \multirow{-1}{*}{\textcolor{white}{\textbf{Non-IC}} } &\multirow{-1}{*}{ \textcolor{white}{ \textbf{IC}}} \\ \hline  \hline
& & \\
 \multirow{-2}{*}{Best connected ($\mathcal{R}_{BC}$)}             &  \multirow{-2}{*}{1.26}    &  \multirow{-2}{*}{-}      \\
 \cellcolor{cyan!20!} & \cellcolor{cyan!20!}    &\cellcolor{cyan!20!}  \\
\multirow{-2}{*}{\cellcolor{cyan!20!} Location Aware $L=2.56/\lambda^\prime$ ($\mathcal{R}_{LA}$)}  & \multirow{-2}{*}{\cellcolor{cyan!20!}1.18} & \multirow{-2}{*}{\cellcolor{cyan!20!}1.22}  \\
  &  &  \\
 \multirow{-2}{*}{ Location Aware $L=0.77/\lambda^\prime$ ($\mathcal{R}_{LA}$)}           &  \multirow{-2}{*}{1.01}     & \multirow{-2}{*}{1.08}     \\  \hline
\end{tabular}
}
\vspace{3mm}
\label{tab3}
\end{table}

\section{Conclusion}

This paper sheds light on the negative impact of cellular network densification due to the imposed excessive handover rate. Particularly, the paper studies the average throughput decay with user velocity in dense cellular environments. To this end, the paper proposes simple yet effective HO management schemes via topology aware HO skipping. The proposed schemes take user location and/or cell-size into account to make HO decisions, thus avoiding unnecessary HOs along the user trajectory. The effectiveness of the proposed schemes are validated in two network scenarios, namely, a PPP single tier cellular network and PCP two tier cellular network. When compared to the conventional best RSS based connected strategy, the proposed skipping models show up to $47\%$ gains in the average throughput over the user velocity ranging from 30 km/h to 240 km/h at BS intensity of 50 BS/km$^2$. Higher gains are expected at higher BSs intensities.




\ifCLASSOPTIONcaptionsoff
  \newpage
\fi

\bibliographystyle{IEEEtran}
\bibliography{IEEEaccess}
\vfill

\end{document}